\documentclass[3p,times,twocolumn]{elsarticle}

\usepackage{ecrc}

\volume{00}

\firstpage{1}

\journalname{Nuclear Physics B Proceedings Supplement}

\runauth{}

\jid{nuphbp}

\jnltitlelogo{Nuclear Physics B Proceedings Supplement}

\usepackage{amssymb}

\usepackage[figuresright]{rotating}

\begin{document}

\begin{frontmatter}



\dochead{}

\title{Optimization of neutrino fluxes for future long baseline neutrino oscillation experiments}


\author[cern]{M. Calviani}
\author[eth]{S. di Luise}
\author[ipnl]{V. Galymov}
\author[cern]{P. Velten}

\address[cern]{CERN, Geneva, Switzerland}
\address[eth]{ETH Z\"urich, Institue for Particle Physics, Switzerland}
\address[ipnl]{Institut de Physique Nucl\'eaire de Lyon, Villeurbanne, France}

\begin{abstract}

One of the main goals of the Long Baseline Neutrino Oscillation experiment (LBNO) experiment is to study the L/E behaviour of the electron
neutrino appearance probability in order to determine the unknown phase $\delta_{CP}$. In the standard
neutrino 3-flavour mixing paradigm, this parameter encapsulates a possibility of a CP violation in
the lepton sector that in turn could help explain the matter-antimatter asymmetry in the universe.
In LBNO, the measurement of $\delta_{CP}$ would rely on the observation of the electron appearance
probability in a broad energy range covering the 1$^{st}$ and 2$^{nd}$ maxima of the oscillation probability. 
An optimization of the energy spectrum of the neutrino beam is necessary to find the best coverage of the neutrino
energies of interest. This in general is a complex task that requires exploring a large parameter
space describing hadron target and beamline focusing elements. In this paper we will present a
numerical approach of finding a solution to this difficult optimization problem often encountered
in design of modern neutrino beamlines and we will show the improved LBNO sensitivity to the
presence of the leptonic CP violation attained after the neutrino beam optimization.
\end{abstract}

\end{frontmatter}


LBNO will utilize a neutrino beam conventionally produced with a high intensity proton beam impinging on a target. 
The proton beam will be delivered at the CERN accelerator complex.
Two options are foreseen for the two successive phases of the experiment. In the first phase, an upgraded SPS will deliver 400 GeV protons at about 700 kW beam power. The expected integrated yearly exposure is about $1.0\times 10^{20}$ protons on target (POT).
In the second phase, a primary beam is foreseen to be provided by a high power PS (HPPS) facility which will deliver a 50 GeV 2 MW proton beam and an integrated yearly exposure of about $3.5\times 10^{21}$ POT.

\begin{figure}[h]
\center
\includegraphics[width=0.5\textwidth]{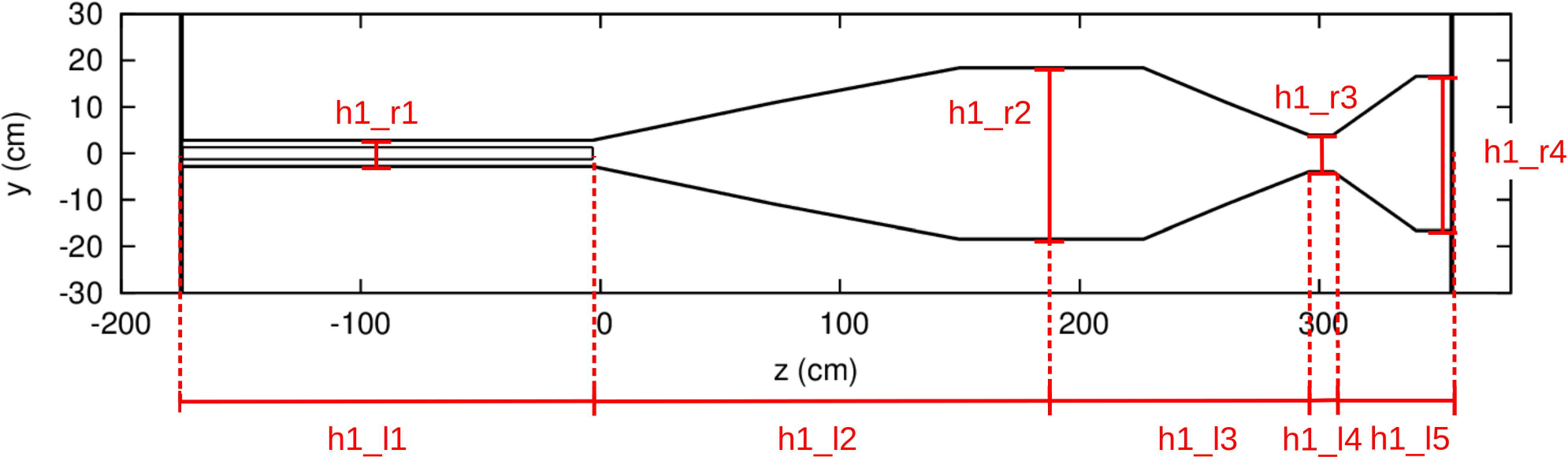}
\includegraphics[width=0.48\textwidth]{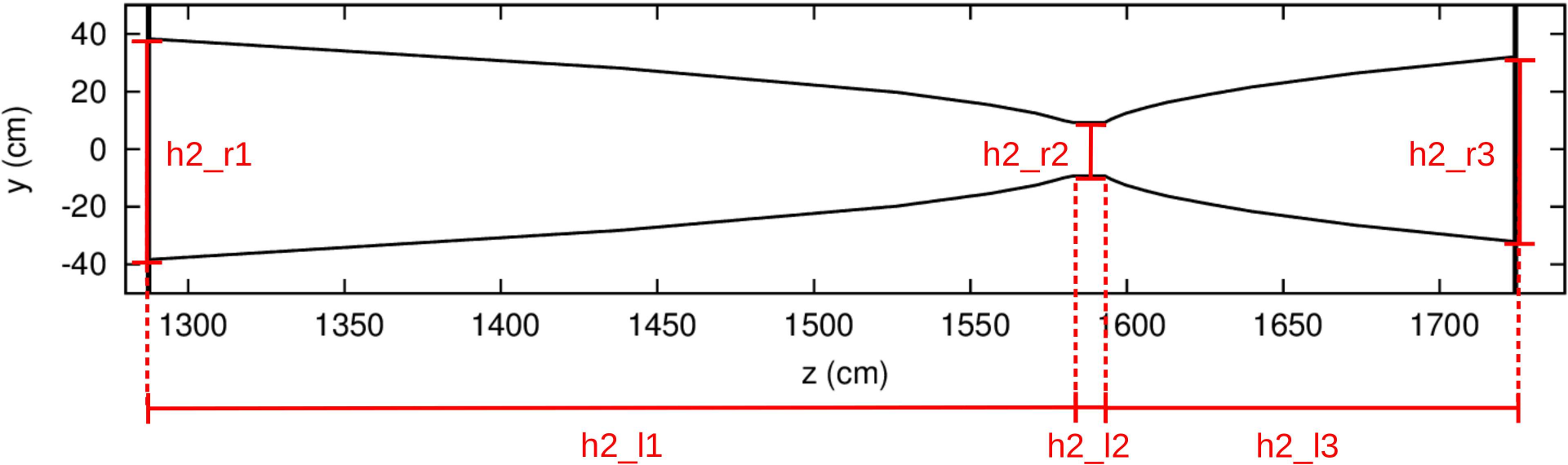}
\caption{Design layout of the 1$^{st}$ horn and target (top) and 2$^{nd}$ horn (bottom).}
\label{fig:beam_par_description}
\end{figure}

\begin{figure}[h]
\center
\includegraphics[width=0.23\textwidth]{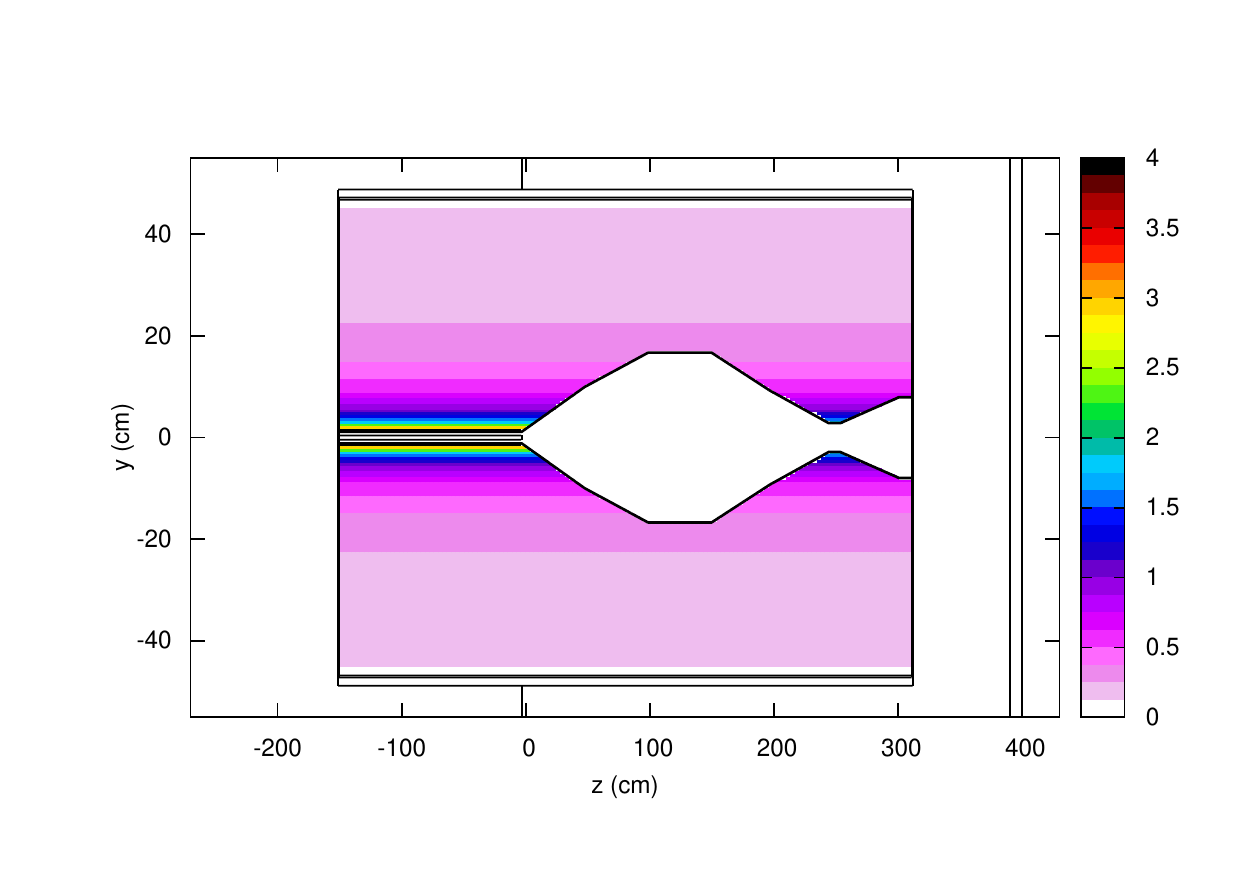}
\includegraphics[width=0.23\textwidth]{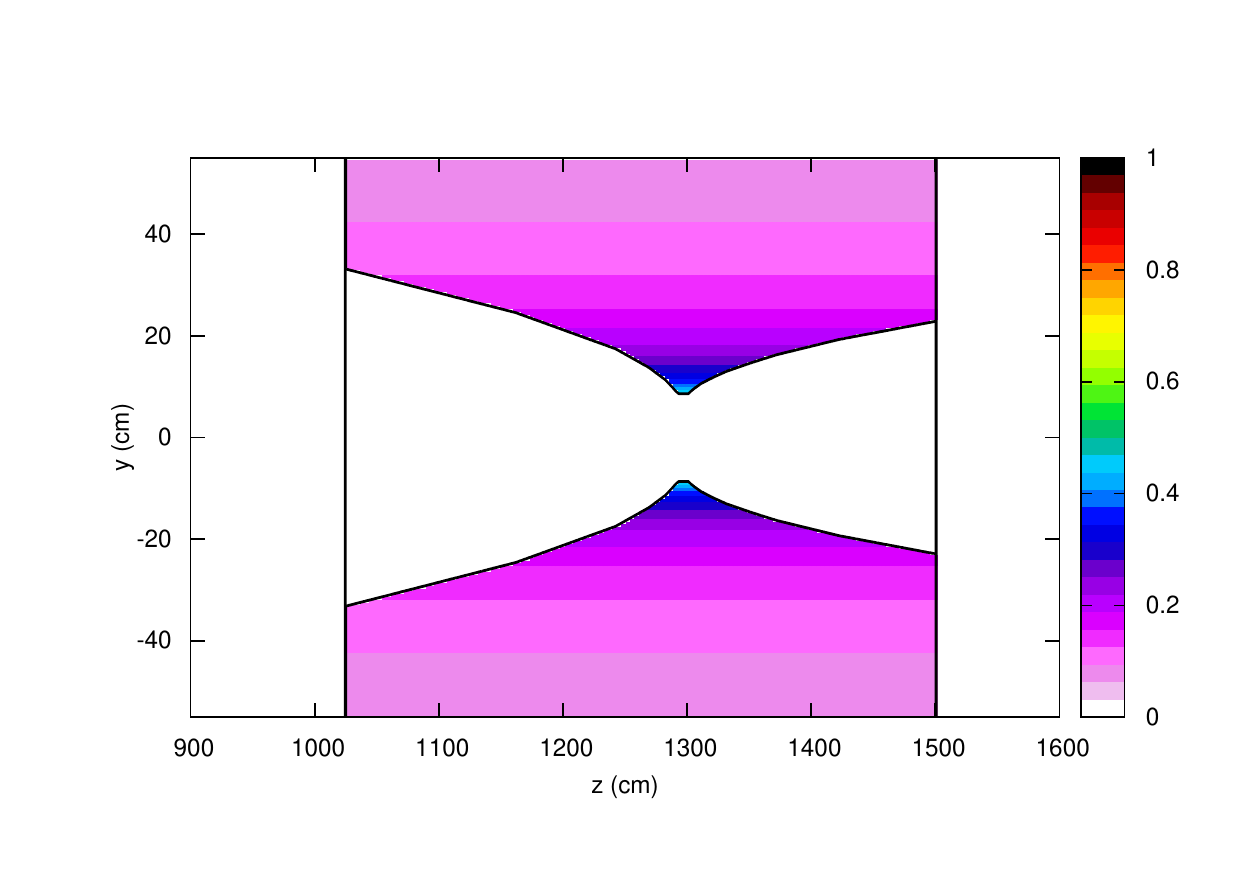}
\includegraphics[width=0.23\textwidth]{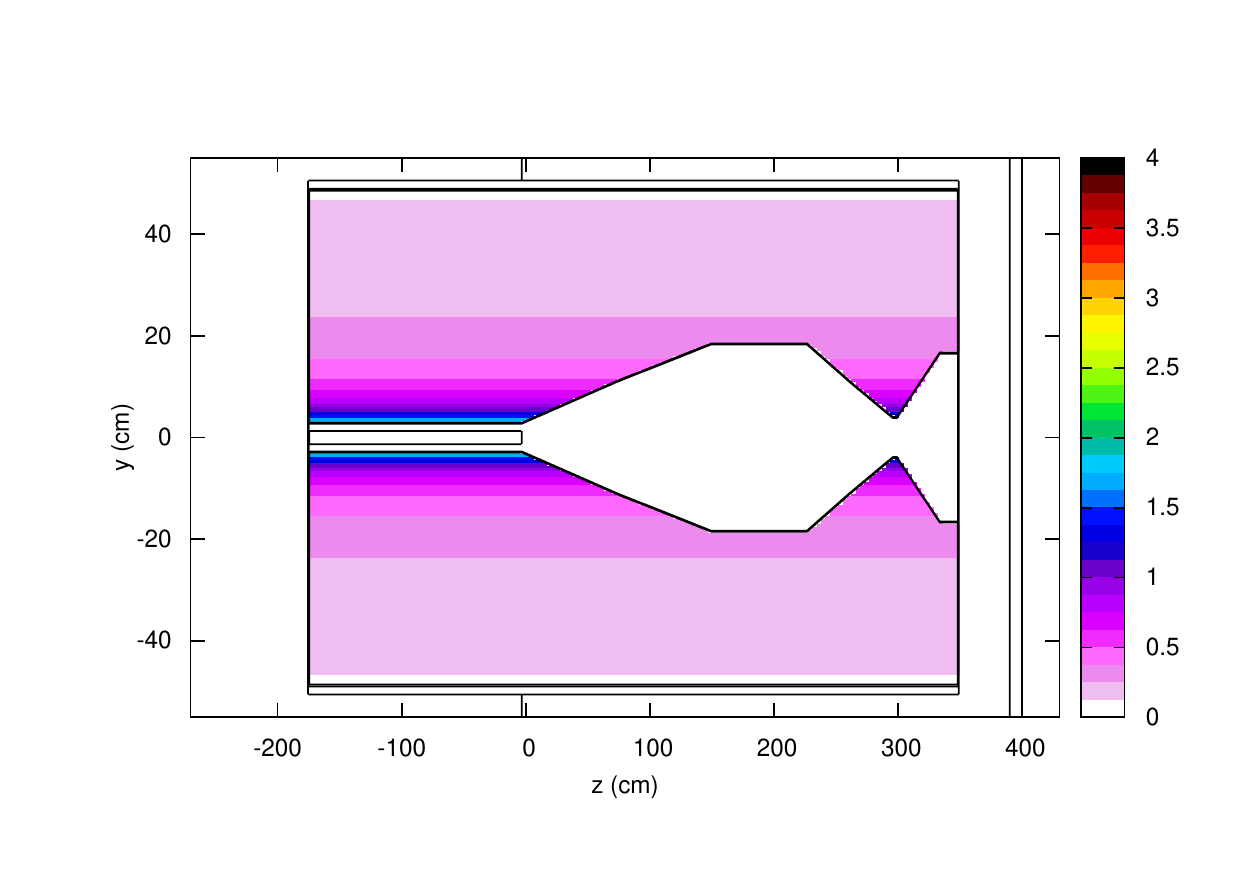}
\includegraphics[width=0.23\textwidth]{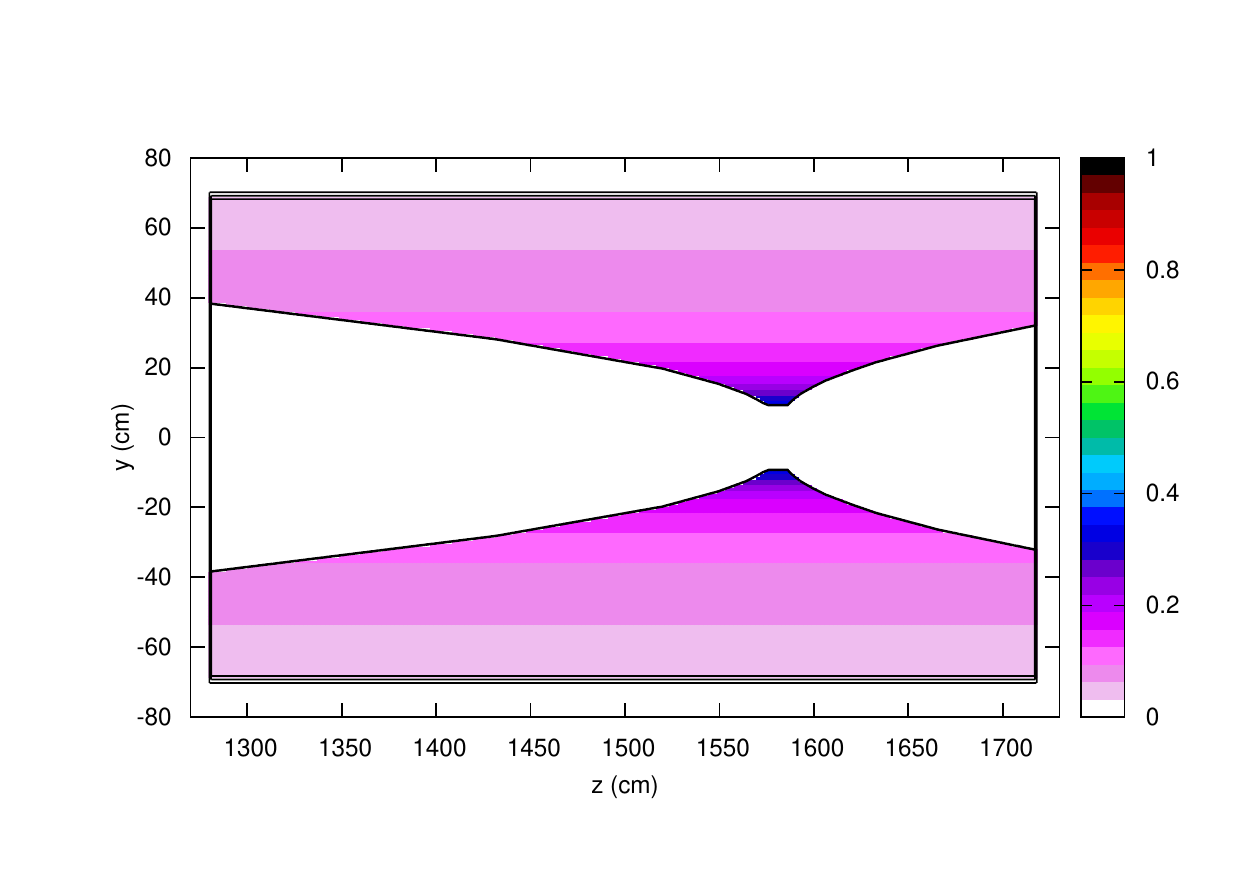}
\caption{View of the horn and target (left) and reflector (right) corresponding to the selected SPS GLB (panels above) and HPPS LE (panels below) configurations. The magnetic field strengths correspond to circulating currents of 281 kA (289 kA) and 198 kA (187 kA) for the 1$^{st}$ horn and 2$^{nd}$ horn respectively and SPS (HPPS) beam option. }
\label{fig:optimbeam_layout}
\end{figure}

A model of the neutrino beamline has been developed in FLUKA\cite{FLUKA} for the calculation of the neutrino flux. In the model, the hadron production target is described as solid graphite cylinder. The focusing optics consists of two aluminum horns. 
The target is fully inserted into the first horn in order to maximize the collection of the low energy pions which contribute to the neutrino flux below 2 GeV (around the 2$^{nd}$ oscillation maximum). At this stage no support system for the target and horns has been modeled and the components are simply placed in an empty environment representing the target station hall. 
The decay tunnel, modeled as a cylinder 300 m long and 3 m in diameter is located 30 m downstream of the target. A hadron beam stop (beam dump) is placed at the end of the decay volume. 
The geometry of the 1$^{st}$ horn adopted in this study differs significantly from simple parabolic horn considered in the LBNO expression of interest \cite{LBNO-EOI}. 
The new design aims to improve collection of the low energy secondaries that exit the target at steep angles above 100 mrad and could not be efficiently bent in the previous design. 
The geometry of the 1$^{st}$ horn (Fig.~\ref{fig:beam_par_description}) is described in terms of ten parameters that define the structure of its inner conductor and the overall dimensions. 
The upstream part of the horn functions as a collector, which aims to minimize the divergence of the beam of secondaries. 
The downstream part, with its elliptical inner conductor shape, attempts to focus particles along the beam axis. 
The 2$^{nd}$ horn or reflector (Fig.~\ref{fig:beam_par_description}) has a more typical inner conductor shape that consists of two ellipsoidal sections. Its geometry is described in terms of seven parameters.
An additional set of parameters defining the target shape, the relative horns position and the circulating horn current completes the
full description of the focusing system.
%
%
%
%
%
%
%
\begin{figure}[h]
\center
\includegraphics[width=0.5\textwidth]{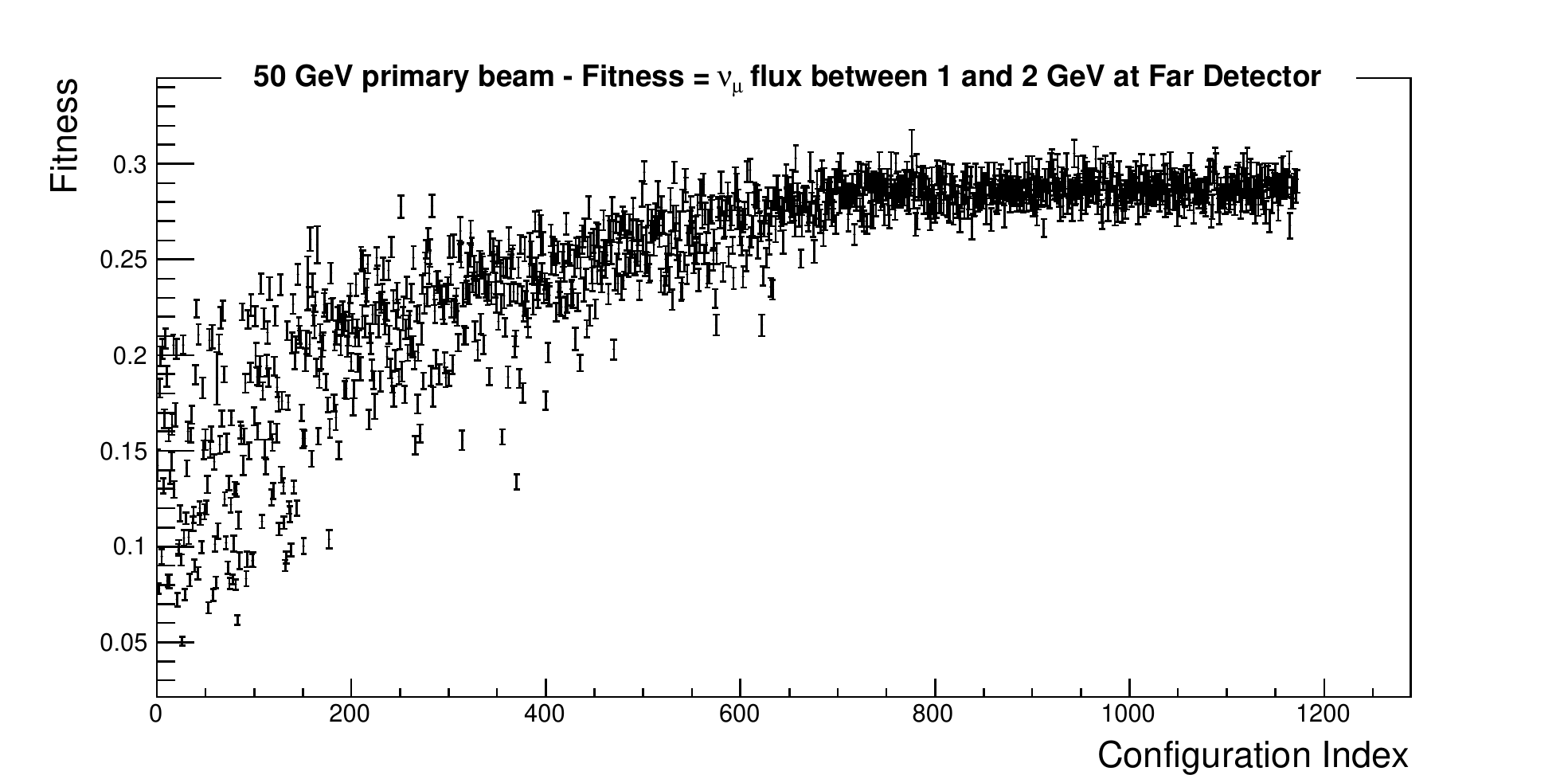}
\caption{Evolution of the fitness parameter during LE optimization of the HPPS-based neutrino beam. The \textit{x} represents a unique index value of a given beamline configuration.}
\label{fig:lefitnessevol}
\end{figure}
\begin{figure}[h]
\begin{center}
\includegraphics[width=0.43\textwidth]{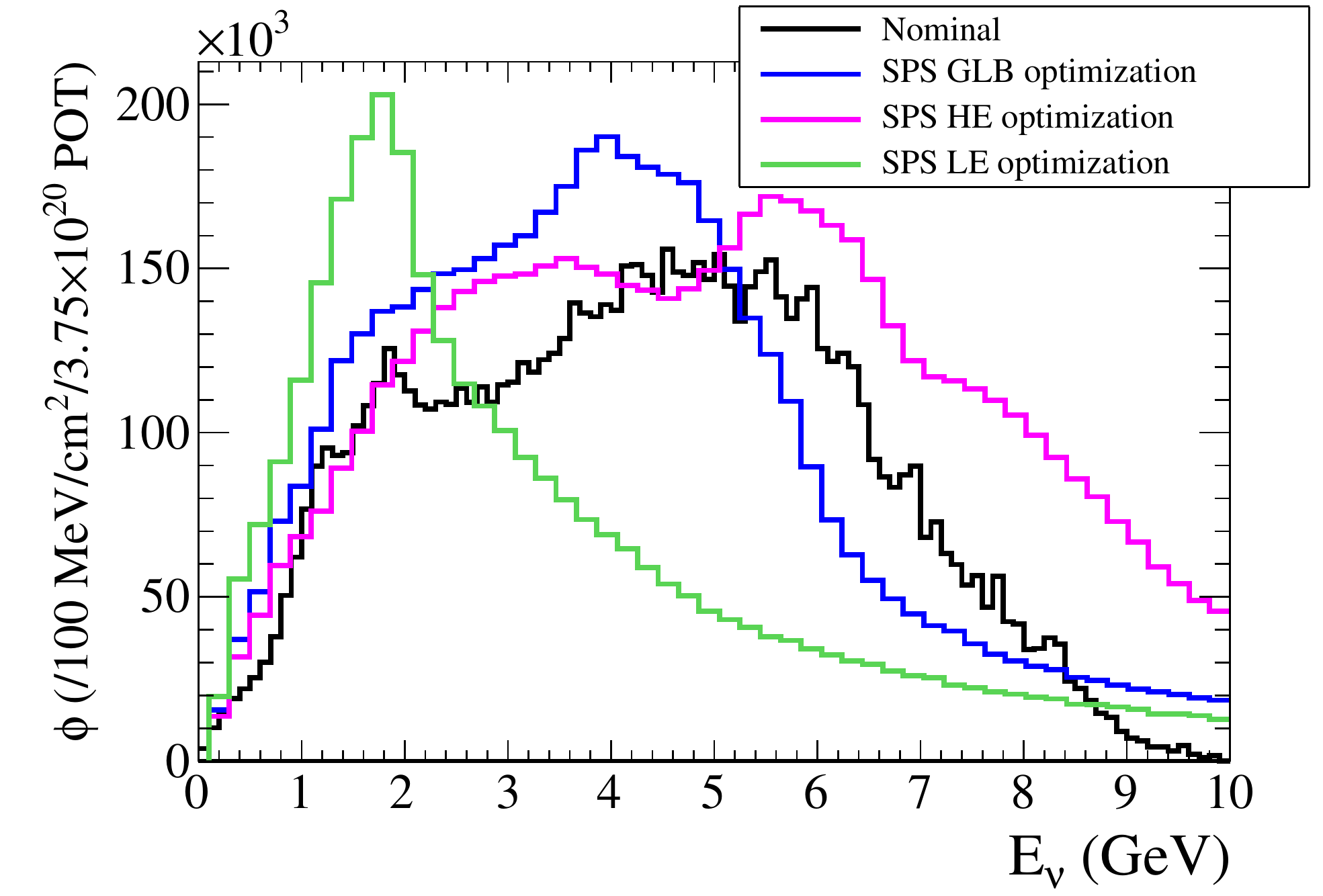}
\includegraphics[width=0.43\textwidth]{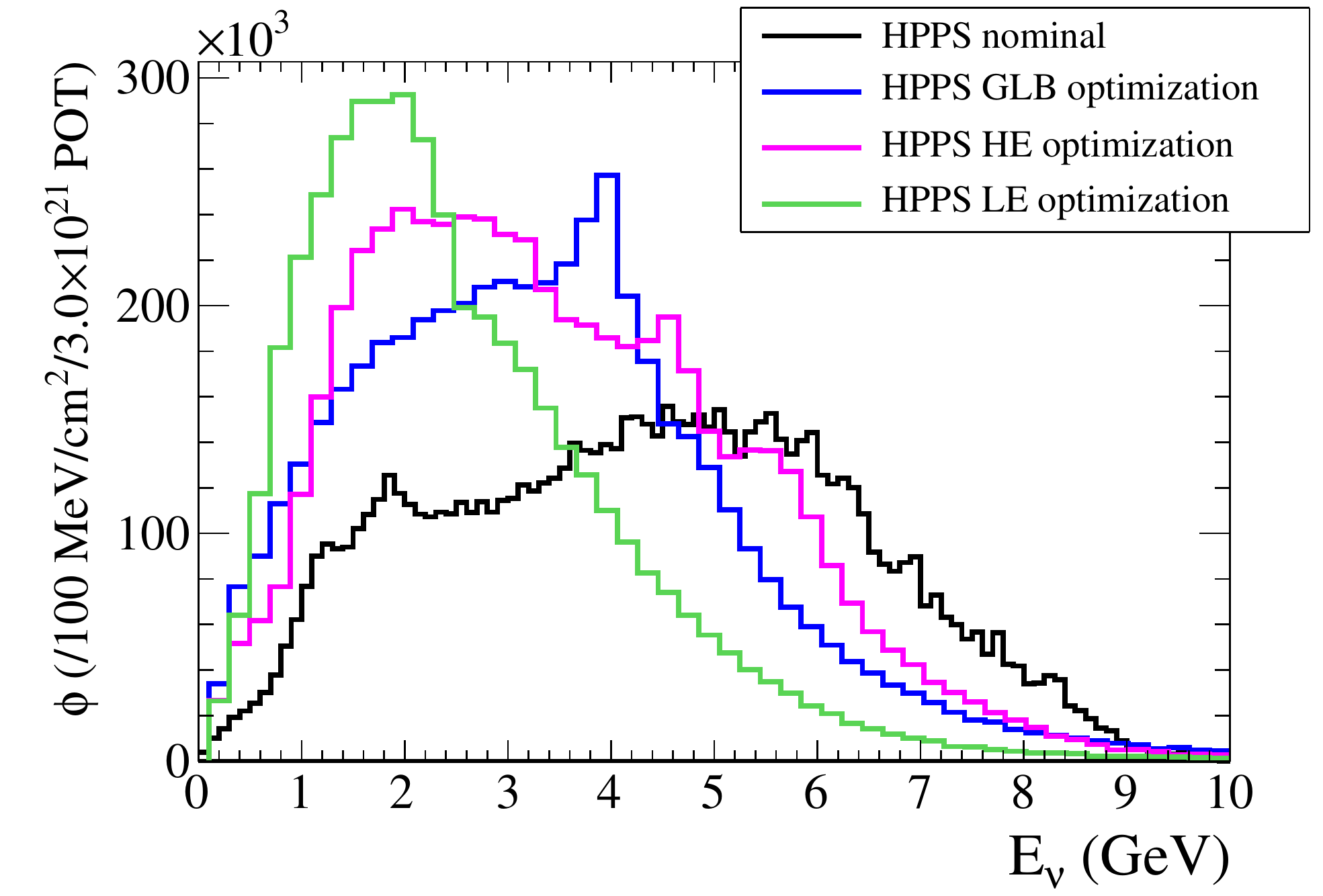}
\caption{Energy spectra of neutrino fluxes for different optimizations for both the SPS (top) and HPPS (bottom) proton beam options. The nominal LBNO flux from \cite{LBNO-EOI} is also shown.}
\label{fig:cern_beams_optim}
\end{center}
\end{figure}
We use genetic algorithm, implemented in DEAP toolkit \cite{DEAP}, to find the optimal values for the parameters. 
The genetic algorithm is an heuristic search that mimics the process of natural selection in order to generate useful solutions to optimization and search problems. A population of candidate solutions, called individuals (the different beamline configurations), is evolved towards a better solution. Each candidate solution has a set of properties, called chromosomes (the parameters of the model). For each generation, a global quantity, the fitness, 
is evaluated for each indiviadual and the best performing individuals are selected to randomly recombine and mutate their chromosomes to produce new individuals. 
This iterative process is carried out until no further improvement in fitness parameter is observed in the population. 
\begin{figure}[h]
\begin{center}
\includegraphics[width=0.43\textwidth]{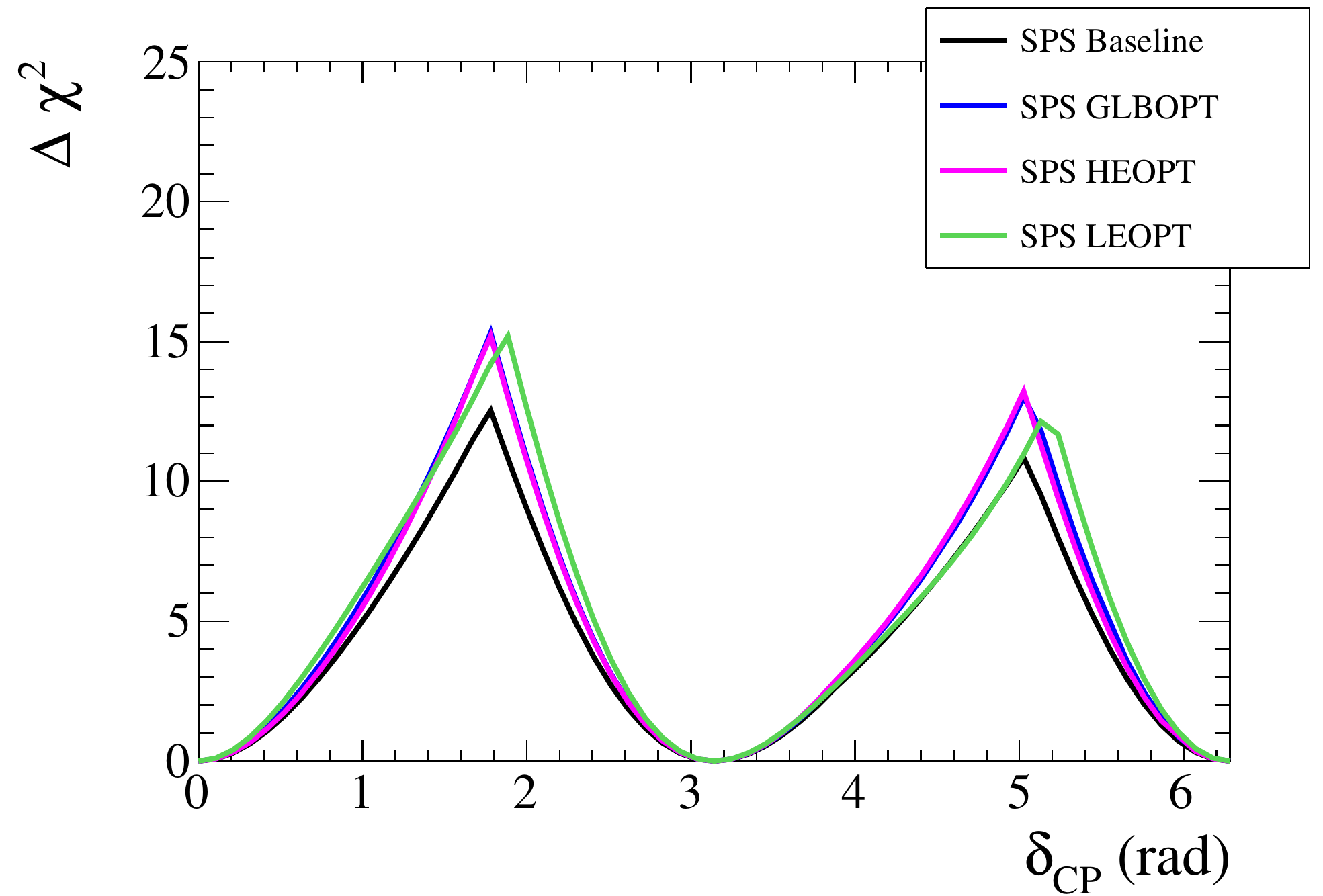}
\includegraphics[width=0.43\textwidth]{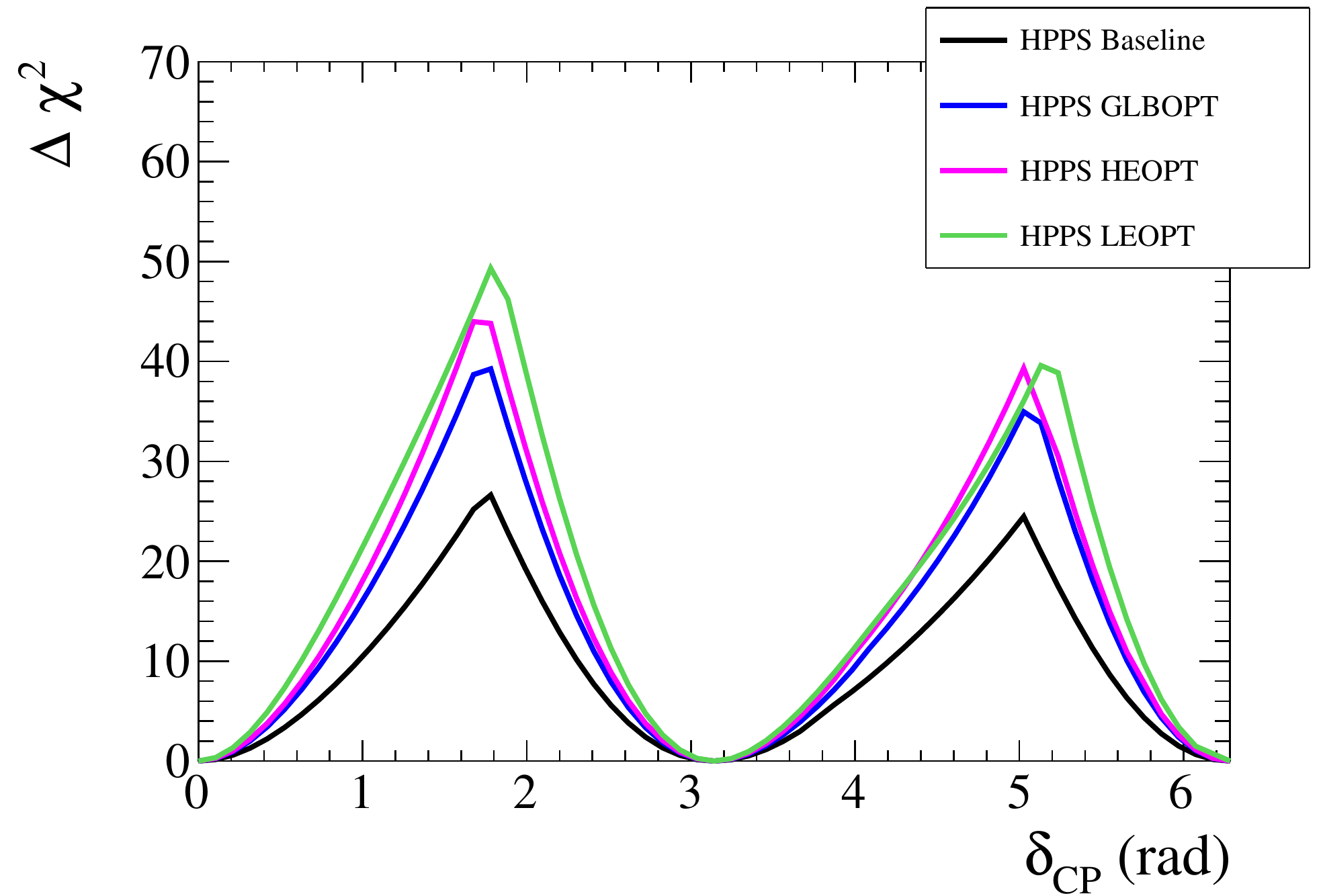}
\caption{Comparison of expected CPV sensitivity for different neutrino flux optimizations for SPS (top) and HPPS (bottom) proton beam options and LBNO20 detector configuration. A total exposure of $15\times 10^{20}$ ($30\times 10^{21}$) POT is taken for SPS (HPPS) beam with 75\% of the running time devoted to the running in the neutrino mode. The value of $\sin^2{\theta_{23}} = 0.5$ is assumed.}
\label{fig:cern_beams_optim_cpv}
\end{center}
\end{figure}
The fitness is what the algorithm will attempt to maximize, therefore is crucial a suitable definition of this variable.
To explore different possible energy windows for the neutrino beam we chose the following three different fitness criteria
\begin{itemize}
\item {\bf High Energy optimization (HE)}: maximization of the integral of $\nu_{\mu}$ flux in a 0-6 GeV energy window. 
In this case, the optimization should generate beam optics configurations producing wide band beams covering both first and second oscillation maxima.
\item {\bf Low Energy optimization (LE)}: maximization of the integral of $\nu_{\mu}$ flux in a 1-2 GeV energy window. 
In this case, beams optics configurations generating a neutrino flux mainly at lower energies around the second oscillation maximum should be obtained.
\item {\bf CPV based optimization using GloBeS (GLB)}: maximization of total $\delta_{CP}$ sensitivity as computed with GloBeS \cite{GLOBES} neglecting all the systematics uncertainties
\end{itemize}
As an example, Fig.~\ref{fig:lefitnessevol} shows the evolution of the fitness during the optimization process. 
Typically, when the algorithm converges, it gives an ensemble of configurations each considered to be optimal from the point of view of the fitness criterion. One has to then choose the best candidate based on considerations related the engineering implementation for a given option.
The energy spectra of the $\nu_\mu$ flux for the best performing configurations in the three optimization schemes are shown in Fig.~\ref{fig:cern_beams_optim} for SPS and HPPS options. As can be seen in the figure, different optimization schemes result in different neutrino energy spectra. To choose the best spectrum for the measurement of $\delta_{CP}$, we process each optimization result through the full LBNO analysis framework 
and calculate the sensitivity to CPV\cite{LBNO,LBNO-ICHEP2014}. The results are shown in Fig.~\ref{fig:cern_beams_optim_cpv} 
In general, each optimized neutrino flux offers a better performance over the nominal LBNO flux with a particularly large improvement achieved by the LE optimization of the HPPS beam. In the case of the SPS beam, no significant difference between GLB and HE optimization are observed in the CPV sensitivity. Both of these beams, however, offer a slightly better performance compared to the results obtained with the SPS LE optimization. The corresponding target-horn layouts for these configurations are shown in Fig.~\ref{fig:optimbeam_layout} for both primary proton beam options.
%
%
%
\nocite{*}
\bibliographystyle{elsarticle-num}
\bibliography{martin}
%
%
%

\end{document}